\documentclass[preprint,
superscriptaddress,
 amsmath,amssymb,
 aps,
 prl
]{revtex4-2}
\usepackage{lineno}
\usepackage{graphicx}
\usepackage{dcolumn}
\usepackage{bm}
\usepackage{units}
\usepackage{float}
\usepackage{amsmath}
 
\usepackage{dblfloatfix}
\usepackage{xcolor}
\usepackage[utf8]{inputenc}
\usepackage[T1]{fontenc}

\usepackage{amsfonts}
\usepackage{booktabs}

\usepackage{hhline}
\usepackage{multirow}
\usepackage{array}

\newcolumntype{P}[1]{>{\centering\arraybackslash}p{#1}}

\usepackage{mathtools}

\begin{document}

\title{Laser-driven electron source suitable for single-shot Gy-scale irradiation of biological cells at dose-rates exceeding $10^{10}$ Gy/s.}

\author{C.A.~McAnespie}
\affiliation{Centre for Light-Matter Interactions,
  School of Mathematics and Physics,
  Queen's University Belfast
 , BT7 1NN, Belfast United Kingdom}

  \author{P. Chaudhary} 
\affiliation{Patrick G. Johnston Centre for Cancer Research,
  Queen's University Belfast
 , BT7 1NN, Belfast United Kingdom}
 
  \author{L. Calvin}
\affiliation{Centre for Light-Matter Interactions,
  School of Mathematics and Physics,
  Queen's University Belfast
 , BT7 1NN, Belfast United Kingdom}
 
\author{M.J.V.~Streeter} 
\affiliation{Centre for Light-Matter Interactions,
  School of Mathematics and Physics,
  Queen's University Belfast
 , BT7 1NN, Belfast United Kingdom}

 \author{G.~Nersysian} 
\affiliation{Centre for Light-Matter Interactions,
  School of Mathematics and Physics,
  Queen's University Belfast
 , BT7 1NN, Belfast United Kingdom}
 
 \author{S. J. McMahon} 
\affiliation{Patrick G. Johnston Centre for Cancer Research, 
  Queen's University Belfast
 , BT7 1NN, Belfast United Kingdom}
 
 \author{K. M. Prise} 
\affiliation{Patrick G. Johnston Centre for Cancer Research,
  Queen's University Belfast
 , BT7 1NN, Belfast United Kingdom}
 
 \author{G. Sarri} 
\affiliation{Centre for Light-Matter Interactions,
  School of Mathematics and Physics,
  Queen's University Belfast
 , BT7 1NN, Belfast United Kingdom}

\begin{abstract}

We report on the first systematic characterisation of a tuneable laser-driven electron source capable of delivering Gy-scale doses in a duration of 10 - 20 ps, thus reaching unprecedented dose rates in the range of $10^{10} - 10^{12}$ Gy/s.  
Detailed characterisation of the source indicates, in agreement with Monte-Carlo simulations, single-shot delivery of multi-Gy doses per pulse over cm-scale areas, with a high degree of spatial uniformity. The results reported here confirm that a laser-driven source of this kind can be used for systematic studies of the response of biological cells to picosecond-scale radiation at ultra-high dose rates.
\end{abstract}

\maketitle

\newpage

\section{INTRODUCTION}

\begin{figure}[b!]
    \centering
    \includegraphics[width=0.7\textwidth]{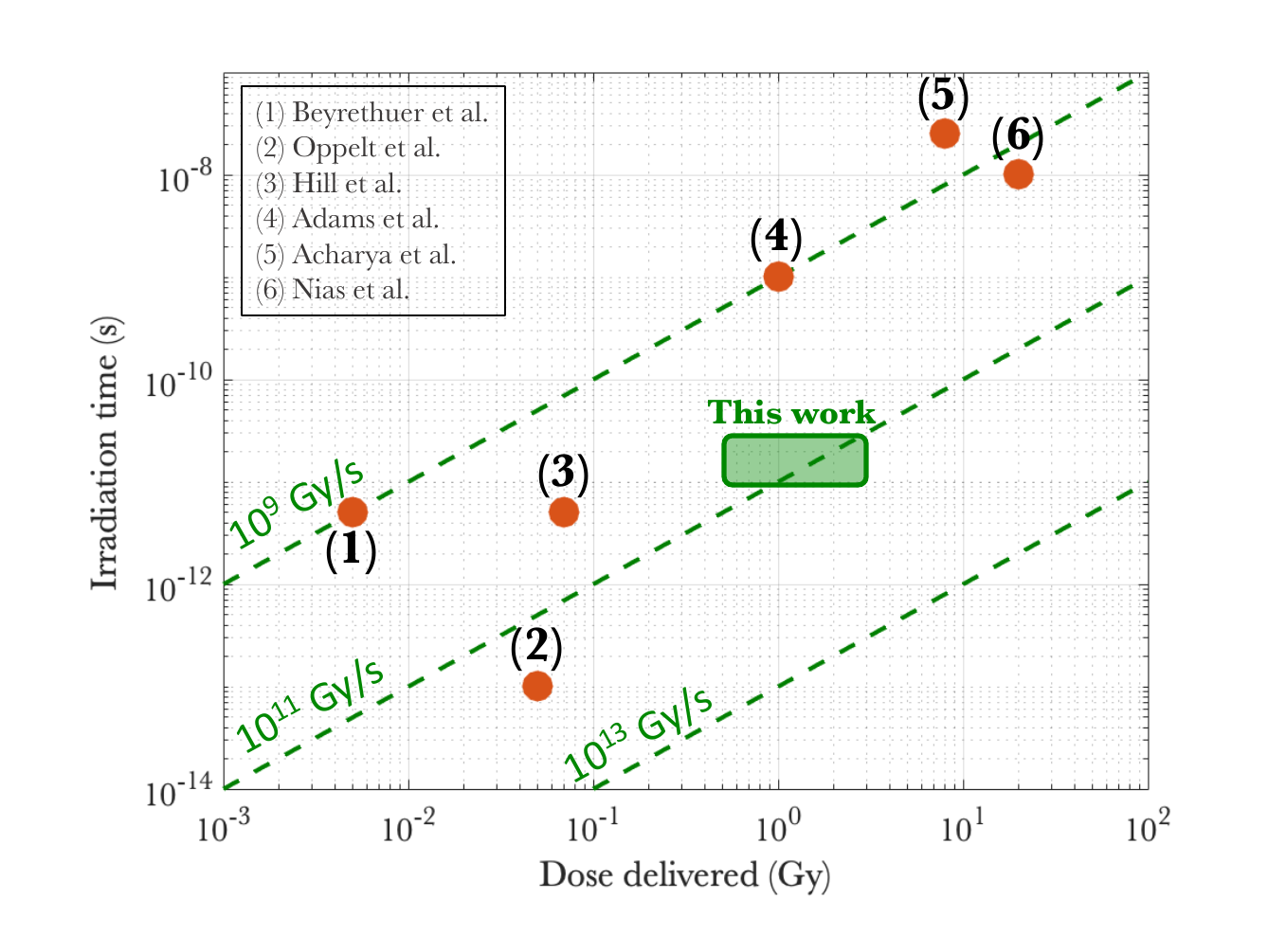}
    \caption{Single-shot dose and irradiation times accessed in this study (green rectangle) compared with experimental work reported in the literature for electron and photon irradiations (orange dots). Dose-rates isocurves are shown for comparison (green dashed lines).}
    \label{fig:previous}
\end{figure}

High dose-rate delivery of radiation to biological cells has been gathering significant attention from the research community with preliminary indications that this technique, generally referred to as FLASH radiotheprapy \cite{harrington2019ultrahigh}, might reduce normal tissue toxicity while maintaining tumour control. Empirical evidence using electrons \cite{schuler2022ultra, schuler2017experimental}, protons \cite{hughes2020flash}, and photons \cite{durante2018faster}, suggests that the sparing of healthy tissue might be linked to radiation-induced oxygen depletion \cite{brown2004exploiting},  even though this theory has been recently questioned \cite{jansen2021does}, further highlighting the need to further understand dose-rate effects before a possible clinical implementation.

While FLASH irradiation delivery usually involves dose-rates in the range of 10s to 100s of Gy/s, an alternative avenue of research has been identified in monitoring the effect of further increasing the dose-rate (\unit[$\gg$100]{Gy/s}), to test possible theories for this sparing effect and identify possible dose-rates at which these effects might be triggered or reach saturation. High-power lasers are ideal tools to study this area of radiobiology since they can now provide radiation sources with unique characteristics, including intrinsic pulse durations ranging from nanoseconds down to tens of femtoseconds \cite{bin2012laser, chaudhary2023cellular, labate2020toward, kokurewicz2021experimental, mcanespie2022high}.  

Proof-of-principle applications of laser-driven sources to radiobiological studies have been reported using both high (i.e., protons and ions) and low (i.e., electrons and photons) linear-energy-transfer (LET) particles \cite{chaudhary2021radiobiology,  kokurewicz2021experimental}. 
However, these experimental studies are still scarce, mainly due to the experimental difficulty in delivering reproducible Gy-scale irradiations with an ultra-short duration. 
To the best of our knowledge, irradiation using laser-driven sources has either been performed as a single-shot Gy-scale irradiation at the nanosecond level \cite{adams1980time, Nias, Acharya, doria2012biological, chaudhary2023cellular}, or as fractionated delivery with single bursts at picosecond or femtosecond level \cite{favaudon2019time, oppelt2015comparison,Hill,Beyrethuer} (see Fig. \ref{fig:previous}). Thus far, no statistically significant deviations in biological endpoints from irradiations at conventional dose rates have been observed. This can be understood by considering that fractionated deliveries, while comprising ultra-short bursts, can still only reach Gy-scale doses over minutes, still resulting in average dose-rates of the order of Gy/s. 


Here, we present an experimental study demonstrating the possibility of delivering single-shot Gy-scale doses over cm-scale areas and with a duration of 10 - 20 ps, reaching unprecedented dose-rates that can be tuned in the range $10^{10}$ - $10^{12}$ Gy/s (see Fig. \ref{fig:previous}). This has been achieved by using MeV-scale electron beams generated during the interaction of a relativistically intense laser pulse with a solid target. The dose properties were monitored, on-shot, with calibrated EBT3 radiochromic films and scintillator screens. The results confirm that an electron source of this kind would be suited to perform radiobiological studies of cellular response to picosecond-scale radiation in a new regime of ultra-high dose rates.


\section{Experimental setup}
\begin{figure}[t!]
    \centering
    \includegraphics[width=0.7\textwidth]{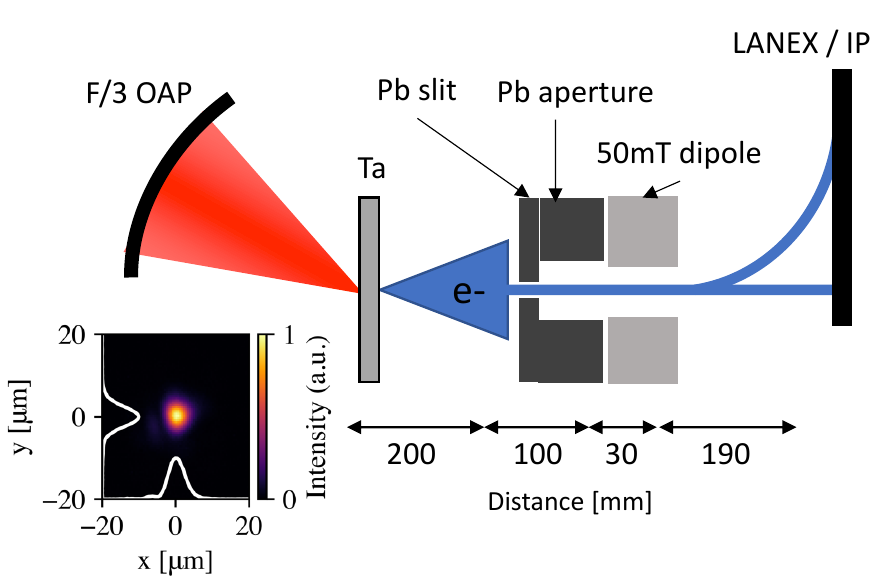}
    \caption{Top-view sketch of the experimental set up. The inset shows the laser focal spot measured at the focal plane.}
    \label{fig:set_up}
\end{figure}

The experiment (sketched in Fig. \ref{fig:set_up}) was performed using the TARANIS laser facility at Queen's University Belfast \cite{dzelzainis2010taranis}. TARANIS is a hybrid Ti:Sapphire - Nd:Glass laser system, delivering \unit[(7.8 $\pm$ 0.3)]{J} in \unit[(0.8 $\pm$ 0.1)]{ps} at a central wavelength of $\lambda=$\unit[1.053]{$\mu$m}. 
The laser was focused using a F/3 off-axis parabola (OAP), down to a focal spot with FWHM of $w_x$ = \unit[(5.2 $\pm$ 0.1)]{$\mu$m} and $w_y$ = \unit[(6.5 $\pm$ 0.1)]{$\mu$m}. 
The resultant peak intensity of the laser pulse on target was I = \unit[(1.7 $\pm$ 0.3) $\times 10^{19}$]{W/cm$^2$}. The laser was incident onto a \unit[50]{$\mu$m} Tantalum foil at an angle of incidence of \unit[30]{\textdegree}. The slightly elliptical shape of the laser focal spot is due to the non-normal incidence on target.

The electron beam properties were measured using a set up as in Fig. \ref{fig:set_up}. After the solid target, a magnetic spectrometer was set up along the target normal axis. A \unit[3]{mm} thick lead slit (\unit[25]{mm} wide) was placed \unit[20]{cm} from the rear of the target, directly followed by a \unit[30]{mm}, 50mT dipole magnet. Mapping of the magnetic field distribution inside the dipole showed a super-gaussian (index=4) magnetic field with $\sigma$=\unit[18]{mm} and peak magnetic field $B_{max}=\unit[51]{mT}$. This magnetic field distribution has been used for particle tracking, to extract the particles' spectrum.
The deflected particles where detected using an image plate (IP), a LANEX scintillator screen or both, mounted vertically \unit[19]{cm} from the rear of the magnet (see Fig. \ref{fig:set_up}). 

To measure the properties of the dose delivered by the electron beam, the magnet was removed from the electron beam path and a set of Radiochromic Films (RCFs) and LANEX scintillator screens were placed on-axis behind a 20 mm aluminium window, at a variable distance from the tantalum target.

\section{Electron beam characteristics}
\begin{figure}[b!]
    \centering
    \includegraphics[width=0.85\textwidth]{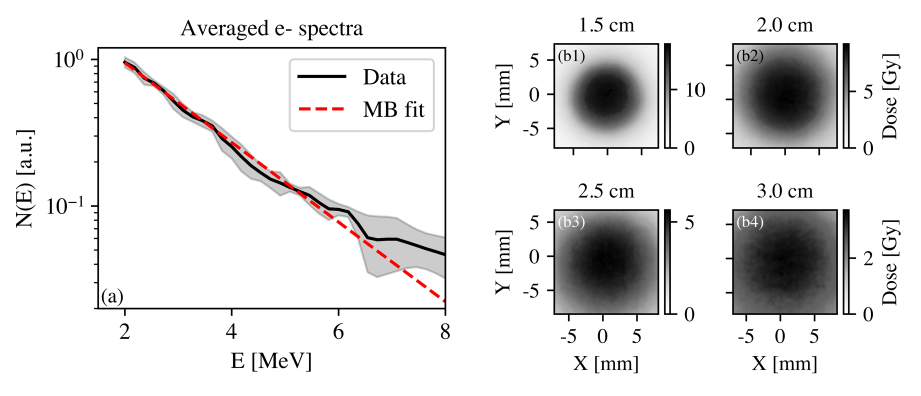}
    \caption{(a) Measured electron spectra: the black solid line indicates the average spectrum over 4 consecutive shots with standard deviation shown as a grey band. The red line shows a Maxwell-Boltzmann fit of the data. (b) Spatial distribution of the dose deposited by the electron beams onto RCFs placed at different distances from the target: \unit[1.5]{cm} (b1), \unit[2]{cm} (b2), \unit[2.5]{cm} (b3), and \unit[3]{cm} (b4). }
    \label{fig:ebeam}
\end{figure}
When an intense laser 
is incident on a solid target, the rising edge of the pulse causes ionisation at the target surface, generating an over-dense plasma. 
The peak intensity of the laser thus interacts with an over-dense plasma rather than an unionised solid. For the intensities of interest here, 
the main laser absorption mechanism is $\vec{J}\times \vec{B}$ heating \cite{wilks1993simulations}, resulting in electron acceleration into the target. This generates a population of super-thermal electrons with a Maxwellian energy distribution with a characteristic temperature that can be estimated as \cite{kruer1985j}:
\begin{equation}
T_{hot} = 511 [\text{KeV}] \left[ \sqrt{1 + \frac{I\lambda^2}{1.34\times10^{18} [W/cm^2 \mu m^2]} } -1\right] 
\end{equation}
where I is the laser intensity in W/cm$^2$ and $\lambda$ is the laser wavelength in micron. For an intensity of \unit[(1.7 $\pm$ 0.3) $\times 10^{19}$]{W/cm$^2$} and a laser wavelength of 1.053 $\mu$m, the resultant hot electron temperature can be estimated as \unit[($1.5\pm0.2$)]{MeV}. Approximately $f\approx10$\% of the laser energy is transferred to the electrons \cite{kruer1985j}, indicating up to $\approx 10^{12}$ electrons per bunch. The pulse duration of these electron bunches at source is of the order of 1.3 times the laser pulse duration \cite{Fuchs_2006} (i.e., $\simeq$ 1 ps) and the cone-angle emission is expected to be of the order of 30$^\circ$ \cite{green2008effect}. 

The electron spectrum obtained in the experiment was well approximated by a Maxwell-Boltzmann distribution, with a characteristic temperature of \unit[$1.6\pm0.2$]{MeV} (Fig \ref{fig:ebeam}(b)). The total number of electrons measured was $1.4\times10^{7}$, in a \unit[$1.8\times$10$^{-5}$] steradian cone, as determined by the lead aperture. 
To measure the electron beam divergence an array of RCFs where placed at incremental distances from the rear of the target surface (see Fig. \ref{fig:ebeam}(b)) and a Gaussian fit was applied to the signal measured in each RCF layer and the standard deviation of the fit recorded in both the $x$ and $y$ direction.
The emission angle was thus measured to be (26.4 $\pm$ 3.7)$^\circ$ and (28.0 $\pm$ 3.3)$^\circ$ in the $x$ and $y$ direction, respectively. 
Therefore, the total number of electrons emitted from the rear side of the target is estimated as $\approx 1.7\times10^{11}$. 
All these experimental values are consistent with the estimates discussed above. 

The electron beam characteristics were found to be stable on  a series of different shots (see, for instance, a comparison between single-shot spectra and their average in Fig. \ref{fig:ebeam} (b)) with a typical shot-to-shot fluctuation in the electron temperature of $<10$\%. 

\section {Detector calibration}
\begin{figure}[b]
    \centering
    \includegraphics[width=\textwidth]{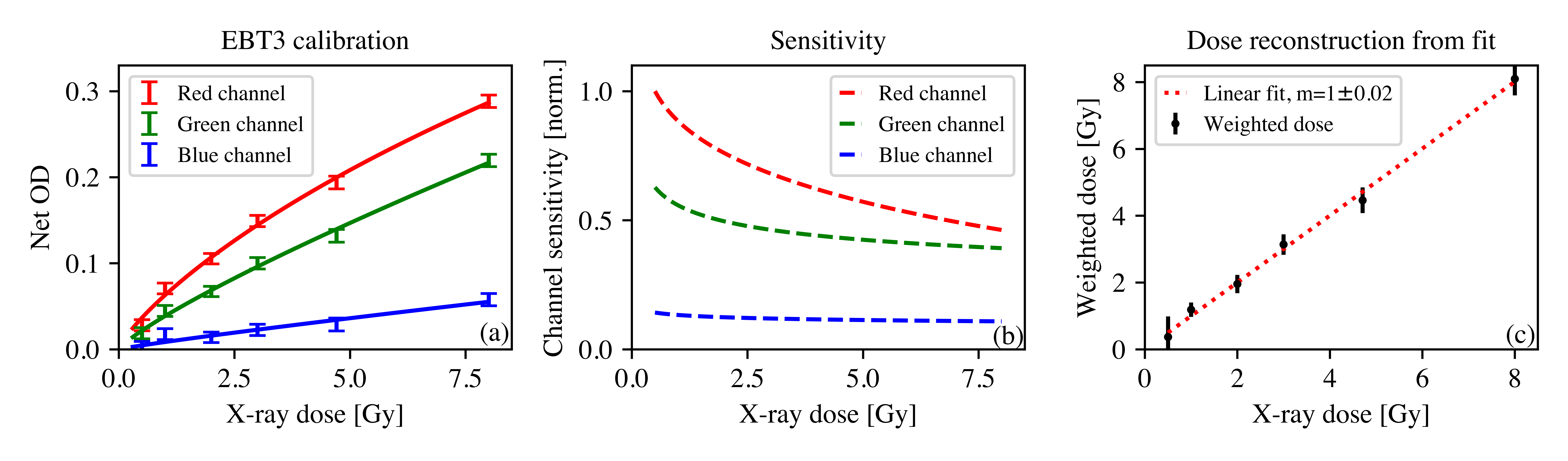}
    \caption{(a) Net optical density response from a range of x-ray doses in each channel (red, green and blue). (b) Sensitivity of each channel as a function of dose deposited. (c): Dose reconstructed from a weighted fit over the 3 channels versus the delivered dose, with a linear fit shown in red (dashed)}
    \label{fig:RCF_calibration}
\end{figure}

For this study, laminated EBT3 radiochromic films were selected due to the recommended dose range (\unit[0.2-10]{Gy}) \cite{campajola2017absolute}. While RCFs are commonly used for dosimetric purposes, it is well-known that their response might vary slightly from batch to batch.  
To confirm the EBT3 dose response in this study, a calibration was then first performed using a \unit[225]{kVp} x-ray source located at PGJCCR at Queens university Belfast, delivering x-ray pulses at a constant dose-rate of \unit[0.49]{Gy/min}. 
The calibration was performed in a dose range of \unit[0.35 - 8]{Gy}, with multiple films irradiated at each dose. 
Each individual RCF can have a different non-irradiated pixel count, so this was measured before each irradiation to allow for accurate dose reconstruction. The dose is reconstructed from the net change in optical density: 
\begin{equation}
 \textrm{OD$_{net}$} = \log_{10}(I/I_0)
 \end{equation}
where 
$I$ and $I_0$ are the pixel values for the exposed and unexposed films, respectively.
The net OD is then compared to the known dose deposited from the x-ray calibration and a fit performed. 
It was found that the dose response in this range is almost linear and can be more accurately reconstructed by the following fit: 
\begin{equation}
 \textrm{Dose} = \textrm{a} \times (\textrm{OD$_{net}$}/(\textrm{b}-\textrm{OD$_{net}$}))^{1/\textrm{c}} 
\end{equation}
where $a$, $b$ and $c$ are fitting parameters. 

The measured net OD in each RBG channel for a known x-ray dose is shown in Fig. \ref{fig:RCF_calibration}(a), with the above fit shown as solid lines. The sensitivity of each channel is shown in Fig. \ref{fig:RCF_calibration}(b) and the reconstructed dose from the three channels as a function of the delivered dose is shown in Fig. \ref{fig:RCF_calibration}(c), showing strong linearity (gradient $m=1.00\pm0.02$).
The error bars on the net OD are a combination of the standard deviation across the central 0.8 by \unit[0.8]{inch} region for both the non-irradiated film and the irradiated film. 
Possible causes for this standard deviation are imperfections in the active layer of the RCF or dust specs on the scanner and RCF film. This uncertainty has been included in the analysis presented in this article.  
It is important to note that it is rather customary to use an averaged value to estimate the non-irradiated $I_0$. However, this might be inaccurate; for example, we found that in the red channel, the non-irradiated $I_0$ (averaged over the central region) varied from 40775 to 41568 between films, resulting in a dose error of $\pm$\unit[0.12]{Gy}. 
Therefore, for most accurate dose reconstruction, each single RCF should be scanned pre- and post- irradiation.  

\section{Dose and dose-rate measurements}
\begin{figure}[b]
    \centering
    \includegraphics[width=\textwidth]{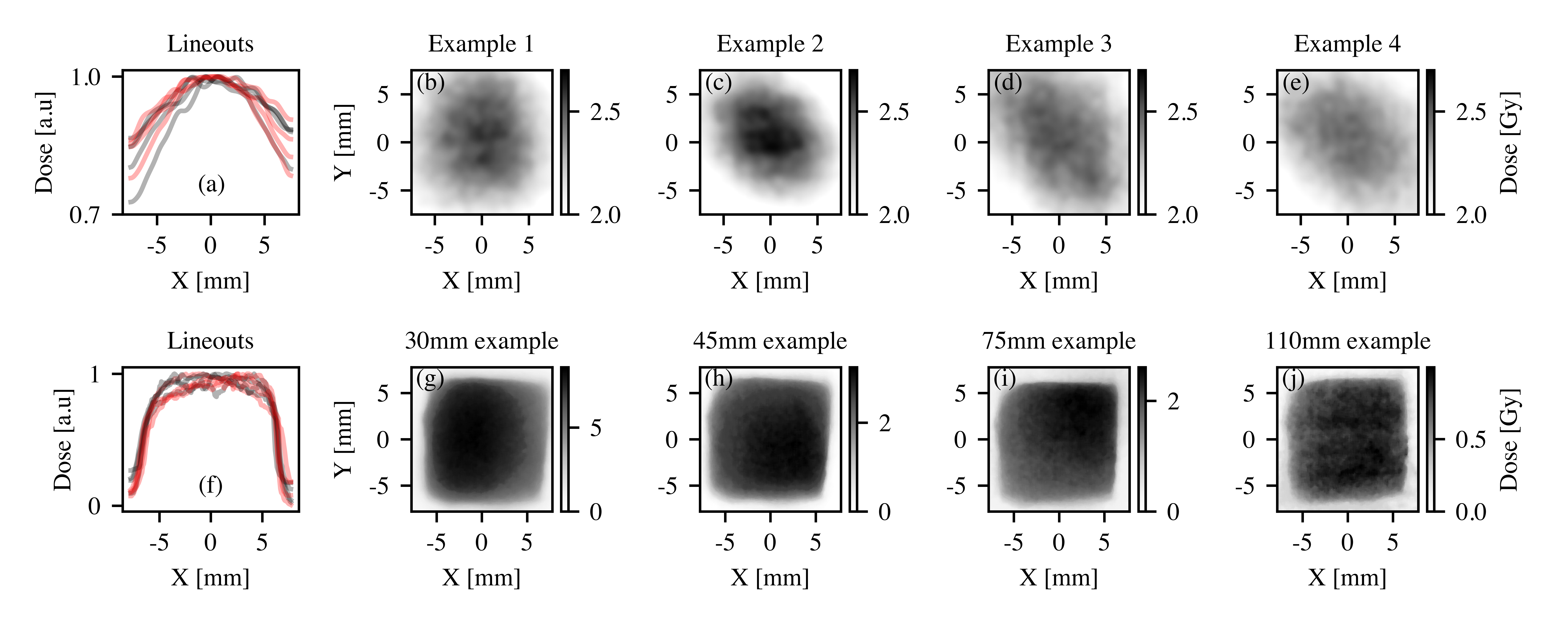}
    \caption{Top row: (a) Normalised lineouts of the dose profile from (b-e) for the vertical (red) and horizontal directions (black). (b-e): Example RCF raw data without lead collimator, at a fixed distance of 55 mm from the solid target. Bottom row: (f) Normalised lineouts of the dose profiles from (g-j). (g-j): Example RCF raw data as a function of distance from the solid target and with the lead collimator in place. Note the different colour-scale for frames in (b-e) and in (g-j).}
    \label{fig:RCF_measurements}
\end{figure}

RCFs were placed in front of the sample to be irradiated for dosimetry in two configurations: fixed distance of 55 mm and without a lead collimator (mimicking requirements for foci formation studies in biological cells) and at variable distances and with a lead collimator (mimicking requirements for cell survival studies as a function of dose delivered). 
Fig. \ref{fig:RCF_measurements}(b-e) shows 4 examples of RCF raw data at a fixed distance of 55 mm, with \ref{fig:RCF_measurements}(a) showing the vertical and horizontal normalised lineouts. 
A Gaussian fit was applied to the lineouts giving a FWHM of \unit[2.98 $\pm$ 0.28]{mm}.
To measure the uniformity of the dose, the coefficient of variance (CoV, defined as standard deviation divided by the mean) was extracted  to provide a relative variation across the central 1$\times$1 cm$^2$ region. For the results shown in Fig. \ref{fig:RCF_measurements}(b-e) the CoV was measured as 3.8, 6.2, 3.8 and \unit[3.4]{\%}, respectively. This area is more than sufficient to irradiate a large number of cells (>1$\times 10^6$), with uniformity in line with other laser-driven radiobiological studies (see, for instance, \cite{oppelt2015comparison, brack2020spectral}). 

To assess the feasibility of performing cell survival studies, the source to sample distance was varied to control the dose delivered to the cell samples. For the four examples shown in Fig. \ref{fig:RCF_measurements}(g-j), the CoV in the central 1$\times$1 cm$^2$ region was found to be 8.2, 7.3, 9.8 and \unit[7.1]{\%}, respectively. Normalised lineouts are shown in Fig. \ref{fig:RCF_measurements}(f), highlighting the presence of the lead aperture. For a distance of 30 mm from the solid target, a maximum dose in excess of 5 Gy was recorded, which progressively decreased down to < 1 Gy at 110 mm from the solid target, while maintaining good spatial uniformity.

\begin{figure}[t!]
    \centering
    \includegraphics[width=0.6\textwidth]{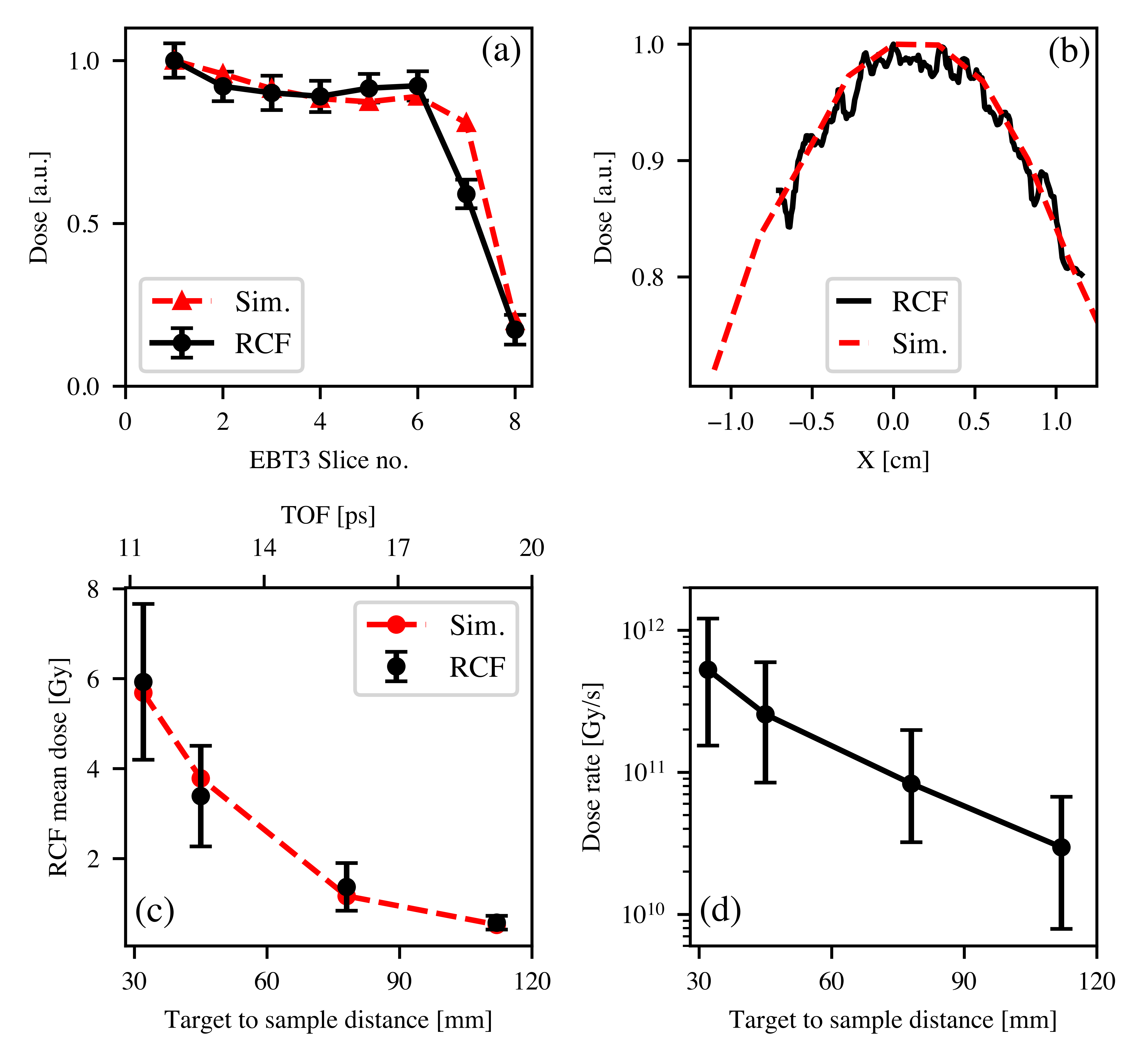}
    \caption{(a) Dose reconstruction on a RCF stack (black solid line) and simulation results for a \unit[1.6]{MeV} spectrum in the same set up (dashed red line). Connecting lines are provided only as a guide for the eye. (b) Example of a lineout of the typical dose deposited in an RCF layer (solid black line) and corresponding simulation result (dashed red line). (c) Mean dose measured as a function of distance from the target (black dots and lower x-axis scale) and $1/e^2$ duration of irradiation (black dots and upper x-axis scale) compared with simulation results (dashed red line). Error bars represent the standard deviation between a minimum of four shots for each data point. (d) Resulting dose-rate as a function of distance from the target. The black line is provided only as a guide for the eye.} \label{fig:RCF_dose}
    \label{fig:RCF_Data_dose_rate}
\end{figure}

The dose-depth profile was experimentally tested by irradiating a stack of RCF and lead filters. 
The stack was placed behind an aluminium window and was made of five RCFs followed by three units each comprising a \unit[25]{$\mu$m} lead foil and  an RCF.
The dose recorded by the stack is shown in Fig. \ref{fig:RCF_dose}(a). Due to its relatively low energy, the electron beam presents a structured dose-depth profile, which rapidly decreases as a function of depth in the stack. 

Both the transverse and dose-depth profiles were simulated using the TOPAS Monte-Carlo code (Geant4 based) \cite{perl2012topas, agostinelli2003geant4, allison2016recent}. The electrons were modeled with a Maxwell-Boltzmann energy distribution with a temperature of \unit[1.6]{MeV}. The source size and divergence were simulated to be 6$\mu$m FWHM and \unit[28]{\textdegree}, respectively. These properties are taken from the experimental measurements of the electron beam properties (see Figs. \ref{fig:ebeam}(b) and (c)). Both the measured depth-dose profile (Fig. \ref{fig:RCF_dose}(a)) and the transverse distribution (Fig. \ref{fig:RCF_dose}(b)) of the dose are well reproduced by the simulations. 

The peak dose deposited  \unit[55]{mm} from the source was \unit[2.4 $\pm$ 0.6]{Gy}. 
Comparing the experimental mean dose to the dose obtained from simulations indicates that this was deposited by a total number of 2$\times10^{11}$ electrons in a \unit[0.84]{str} cone, which is consistent with the number of electrons recorded by the image plate in the experiment (1.7$\times10^{11}$). 

Fig. \ref{fig:RCF_dose}(c) shows the mean dose deposited at the four different positions. As stated before, the divergence of the electron beam implies that progressively lower doses are deposited further away from the target, with a trend that is well reproduced by a $1/r^2$ dependence on distance $r$, ranging from \unit[5.9$\pm$1.7]{Gy} at \unit[30]{mm} down to \unit[0.6$\pm$0.2]{Gy} at \unit[110]{mm}. The measured dose is in good agreement with the simulation results.

The electron beam has a duration at source of the order of 1 ps (i.e., approximately 1.3 times the laser pulse duration \cite{Fuchs_2006}). However, the relatively low energy and broad spectrum of the beam implies that, during propagation, time-of-flight effects are non-negligible, and the beam duration increases over distance. 
Simulation results indicate that the electron beam has a duration of approximately \unit[11]{ps} \unit[30]{mm} away from the target, which further increases to approximately \unit[20]{ps} at \unit[120]{mm} (see Fig. \ref{fig:RCF_dose}(c)). As such, the dose rate at different distances from the target is between $3\times10^{10}$ and \unit[$6\times10^{11}$]{Gy/s} (see Fig. \ref{fig:RCF_dose}(d)). 



It must be noted that other sources of dose could be generated during a laser-plasma interaction of this kind. For example, it is well-known that laser-solid interactions, like the one described here, can generate bunches of energetic protons from the rear surface via target normal sheath acceleration \cite{MacchiRMP}. In similar experimental conditions, a maximum proton energy of \unit[$\leq$10]{MeV} has been obtained using the TARANIS laser (See Ref. \cite{dzelzainis2010taranis} for results and \cite{fuchs2006laser} for scaling laws).
Such proton beams are completely stopped \unit[$\approx$ 6]{mm} into the \unit[20]{mm} aluminium window, thus depositing no dose on the RCFs. 

Another potential contribution to the dose deposited at the cell plane arises from the bremsstrahlung x-rays emitted from the tantalum target. Monte-Carlo simulations show that the electrons propagating through the \unit[50]{$\mu$m} tantalum target generate a bremsstrahlung photon beam with a characteristic energy of $\approx$\unit[60]{keV}, together with an additional photon population generated during the propagation of the electron beam through the aluminium window. 
TOPAS simulations indicate a total dose deposited by x-rays of the order of \unit[2$\times$10$^{-13}$]{Gy} per primary electron, to be compared with \unit[4$\times$10$^{-11}$]{Gy} deposited by each primary electron. The x-ray contribution to the dose deposited thus accounts for $<1$\% of the total dose, and can be neglected.




\section{CONCLUSIONS}


We have presented a systematic study of the properties of the dose delivered by MeV-scale electron populations generated during the interaction of a relativistically intense laser pulse with a solid target. Experimental results, in good agreement with Monte-Carlo simulations, indicate that maximum doses in excess of 5 Gy can be delivered in a single shot over cm-scale areas and with a good degree of spatial uniformity (coefficient of variance consistently below 10\%). Due to the large divergence of the electron beam population, it is possible to seamlessly control the dose delivered by varying the distance of the sample from the solid target, and thus perform studies of survival of biological cells as a function of dose delivered. The electron beam duration is calculated to be between 10 and 20 ps, resulting in dose rates in the range of $10^{10}$ - $10^{12}$ Gy/s. 

The dose properties presented here are well suited to perform systematic studies of radiobiological response of biological cells to picosecond-scale radiation, i.e., in a time-scale comparable to the first onset of physico-chemical mechanisms in the cell, and at unprecedently high dose-rates, allowing for experimental studies of potential non-linear and inter-track effects.  

Furthermore, single-shot doses in excess of \unit[8]{Gy} are in principle achievable by varying the laser and target parameters, and can be used to extend cell survival studies over dose ranges of interest to assess potential FLASH-like effects at these ultra-high dose-rates. For example, preliminary experimental parametric scans of the electron beam properties as a function of target thickness indicates that a 25 $\mu$m-thick tantalum target induces a 5-fold increase in electron number, when compared to the 50 $\mu$m reported here, while maintaining a similar spectral shape and temperature.  Further decreasing the target thickness induces a reduction in electron number, possibly due to the non-ideal temporal contrast of the laser.


\section*{Acknowledgements} 
The authors wish to acknowledge support from EPSRC (grant numbers: EP/V044397/1 and EP/P010059/1). SJM is supported by a UKRI Future Leaders Fellowship, grant number MR/T021721/1. M.J.V.S. acknowledges support from the Royal Society URF-R1221874.  KMP acknowledges support from Brainwaves NI.

\section*{REFERENCES}

\bibliographystyle{apsrev4-1}
\bibliography{main_final.bib}



\newpage

\end{document}